\documentclass[pre,aps,amsmath,amssymb,superscriptaddress]{revtex4}
\usepackage{mathrsfs}
\usepackage{amssymb}
\usepackage{graphicx}
\usepackage{amsmath}
\usepackage{bbm}
\usepackage{graphicx}
\usepackage{dcolumn}
\usepackage{bm}
\newcommand{\bra}[1]{\left(#1\right)}
\newcommand{\brb}[1]{\left[#1\right]}
\newcommand{\brc}[1]{\left \langle#1\right \rangle}
\newcommand{\bre}[1]{\left\{#1\right\}}
\newcommand{\brae}[1]{\left( \left\{#1\right\}\right)}

\newcommand{\be}{\begin{equation}}
\newcommand{\ee}{\end{equation}}
\newcommand{\bea}{\begin{eqnarray}}
\newcommand{\eea}{\end{eqnarray}}

\DeclareMathOperator{\Tr}{Tr}
\begin{document}
\title{Non-backtracking operator for Ising model and its application in attractor neural networks}
	\author{Pan Zhang}
\affiliation{Santa Fe Institute, Santa Fe, New Mexico 87501, USA}
\email{pan@santafe.edu}
\begin{abstract}
	The non-backtracking operator was recently shown to give a redemption for spectral 
	clustering in  sparse graphs.
	In this paper we consider non-backtracking operator for Ising model on a general 
	graph with a general coupling
	distribution by linearizing Belief Propagation algorithm at 
	paramagnetic fixed-point. The spectrum of the operator is studied, the sharp edge of bulk 
    and possible real eigenvalues outside the bulk are computed analytically 
	as a function of couplings and temperature.
	We show the applications of the operator in attractor neural networks.
	At thermodynamic limit, our 
	result recovers the phase boundaries of Hopfield model obtained by replica 
	method. On single instances of Hopfield model, its eigenvectors can be used to
	retrieve all patterns simultaneously. We also give an example on how to control the 
	neural networks, i.e. making network more sparse while keeping patterns stable, 
	using the non-backtracking operator and matrix perturbation theory.
\end{abstract}
	\maketitle
\section{Introduction}
Ising model, which consists of spins arranged in a graph, is a fundamental model in 
statistical physics. It also has 
direct applications in many fields of science including e.g.
physics \cite{simon1993statistical, Mezard1987}, neuron science \cite{Schneidman2006}, 
computer science \cite{Hopfield1982, Ackley1985a} and social science \cite{castellano2009statistical}.
With different distribution of couplings, Ising model can act as model of different systems e.g. 
ferromagnets, spin glasses and neural networks. It gives a simple example of 
phase transitions, which is a phenomenon appearing in different fields of 
science, and it can be seen as a simple model for collaborative behavior of iterating systems.

Exact solution of Ising model only exists in special cases of topologies, e.g. one dimensional and 
two dimensional lattices. On a general graphs that the most applications meet, one usually needs 
approximations, especially with a general distribution of couplings. Commonly used approximations 
in statistical physics includes na\" ive mean-field approximation, TAP approximation
and Kikuchi expansions. Among those approximation, one of the most popular approximation is 
Bethe approximation which assumes independence of
conditional probabilities. Bethe approximation is exact on trees, 
and is found to be a good approximation on (random) sparse systems.
Message passing algorithms based on Bethe approximation, so called Belief Propagation (BP)
algorithms and its variations, have been used widely in many fields \cite{Mezard2009}. 

It has been recently proposed in \cite{Krzakala2013} that in network clustering 
(community detection) problem, 
where the task is to 
cluster nodes of graph into groups, Belief Propagation~\cite{Yedidia2001}, 
known in statistical physics as the cavity method~\cite{Mezard2001}, can be approximated by 
spectral clustering using a matrix called ``non-backtracking'' matrix $B$.
The reason is that in the clustering problem where permutation symmetry holds, BP equation 
always has a ``paramagnetic''
fixed-point where every node takes equal probability in each group.
Linearizing BP equation around this paramagnetic fixed-point results to a linearized
version of BP which is equivalent to spectral clustering using non-backtracking matrix on the graph. 
Spectral clustering using ``non-backtracking'' improves significantly the performance of clustering
over other matrices like random walk matrix and Laplacians, and gives a nature choice of number
of groups in the graph. The reason for the improvement is that non-backtracking matrix of a graph as a 
good property that almost all its eigenvalues are confined by a disk in the complex plane, so the 
eigenvalues corresponding to community structure lay outside the disk and are easily distinguishable.

Since the key that relates BP and 
non-backtracking matrix is the 
``paramagnetic'' fixed-point, it is natural to think that similar for graphs, 
there must be also a non-backtracking matrix for 
Ising model when Ising model has a paramagnetic phase at high temperature without external fields.
So the main purpose of this paper is to study 
the non-backtracking operator to Ising model on a graph with a given 
distribution of couplings at a certain temperature, and study some properties on spectrum of the matrix. 
Unlike non-backtracking matrix in graphs, the non-backtracking matrix of Ising model is a 
function of couplings and temperature. In another sense we can think it as a non-backtracking matrix of 
a weighted graph.  
We also study the application of matrix $C$ with some special distribution of couplings where Ising model
acts like spin glasses or associative memory. We show how to use spectrum of $C$ to 
detect phases and phase transitions in the thermodynamics limit of system as well as in single instance 
of small systems. We will also give an example on how to control the neural network using 
eigenvalues and eigenvector of $C$.

The paper is organized as follows. Section~\ref{sec:B} includes definitions of the 
non-backtracking matrix for Ising model, some properties of its spectrum for general couplings are 
computed. 
In Section~\ref{sec:examples} we give several examples on Ising model with specific coupling distributions
, including ferromagnet, Sherrington-Kirkpatrick model and Viana-Bray model. We show that 
in the thermodynamic limit results obtained using non-backtracking matrix recover
the known results of phase boundaries.
In Section~\ref{sec:hop} we consider Hopfield model as associative memory where some binary patterns are 
memorized. In thermodynamics limit we recover the phase boundaries obtained by replica method, it also 
allows one to determine number of patterns and retrieve all the patterns simultaneously.
On single instances we show 
how to make a given network more sparse while keeping patterns memorized using non-backtracking matrix.
We conclude in Section~\ref{sec:dis}.

\section{Non-backtracking operator for Ising model}\label{sec:B}
As we introduced in the introduction, non-backtracking operator appears naturally in linearizing 
BP of inference of 
stochastic block model \cite{Krzakala2013}, also in linearizing BP for modularity \cite{zhang2014a}.
So we can do similar things: obtaining non-backtracking operator by linearizing BP equation 
of Ising model at paramagnetic fixed point.

We consider Ising model on a graph with $n$ spins, $\bre{\sigma}$ is used to denote a configuration, 
with $\sigma_i\in\bre{-1,+1}$ and $i\in \bre{1,...,n}$.
Energy (Hamiltonian) of system with zero external field is pairwise:
$$E\bra{\bre{\sigma}}=-\sum_{\brc{ij}}J_{ij}\sigma_i\sigma_j.$$
Boltzmann distribution with inverse temperature $\beta$ reads
$$P{\bre{\sigma}}=\frac{1}{Z}e^{-\beta E\brae{\sigma}}.$$
On a single instance of the graph, one can use belief propagation algorithm to study the marginals of the 
Boltzmann distribution.
For Ising model, BP equation reads
\begin{equation}\label{eq:bp}
	\psi_{i\to j}=\frac{1}{Z_{i\to j}}\prod_{k\in \partial i\backslash j}\bra{ 2\sinh(\beta J_{ij})\psi_{j\to i}+e^{-\beta J_{ij}}},
\end{equation}
where $\psi_{i\to j}$ denotes ``cavity'' probability of $\sigma_i$ taking value $+1$ with spin $j$ 
removed from the graph, $\partial i$ denotes neighbors of $i$ and 
\begin{equation}
	Z_{i\to j}=\prod_{k\in \partial i\backslash j}\bra{ 2\sinh(\beta J_{ij})\psi_{j\to i}+e^{-\beta J_{ij}}}+\prod_{k\in \partial i\backslash j}\bra{ 2\sinh(-\beta J_{ij})\psi_{j\to i}+e^{\beta J_{ij}}},
\end{equation}
is the normalization.

When BP equation converges on a graph, one can compute marginals of a spin taking value $+1$ using 
\begin{equation}
	\psi_{i}=\frac{1}{Z_i}\prod_{k\in \partial i}\bra{ 2\sinh(\beta J_{ij})\psi_{j\to i}+e^{-\beta J_{ij}}},
\end{equation}
where again $Z_i$ is the normalization.

It is easy to see that $\psi_{i\to j}=\psi_{i}=\frac{1}{2}$ 
for all cavity messages and all marginals is always a fixed point of BP. 
We call this fixed point ``paramagnetic'' fixed-point.
We can write BP messages as deviations from this paramagnetic fixed point:
$$\psi_{i\to j}=\frac{1}{2}+\Delta_{i\to j},$$ with
\begin{equation}\left.\Delta_{i\to j}=\sum_{k\in\partial i\backslash j}\frac{\partial \psi_{i\to j}}
	{\partial \psi_{k\to i}}\right |_{\psi=\frac{1}{2}} \Delta_{k\to i}, \label{eq:lbp} \end{equation}
It is easy to check that derivatives of BP messages evaluated at paramagnetic fixed-point reads
\begin{equation}
	\left.\frac{\partial \psi_{i\to j}}{\partial \psi_{k\to i}}\right |_{\psi=\frac{1}{2}}=\tanh(\beta J_{ik}).
\end{equation}
Equation \eqref{eq:lbp} is a linearized version of BP equation \eqref{eq:bp}, 
and it is equivalent to the eigenvector problem of 
an operator that we call non-backtracking for Ising model:
\begin{equation}
	C_{i\to j,k\to l}=\delta_{l,i}(1-\delta_{k,j})\tanh(\beta J_{lk}),
\end{equation}
which is defined on the directed edges of the graph, with weight $\tanh(\beta J_{ij})$ on each 
directed edge. If we use 
$m$ to denote number of edges in the graph, then the size of $C$ is $2m\times 2m$.
Note that in right hand side of last equation, $\delta_{i,l}(1-\delta_{k,j})$ is 
the non-backtracking operator used in \cite{Krzakala2013} for clustering problem. In 
discussion of \cite{Krzakala2013} authors suggested a weighted non-backtracking matrix, 
and our matrix $C$ can be seen as such weighted non-backtracking matrix, with weight $\tanh(\beta J_{ij})$
on each edge.

Following the technique used in \cite{Krzakala2013},
the edge of bulk of eigenvalues of $C$ can be computed by the following inequality,
\begin{equation}\label{eq:ineq}
	\sum_{a=1}^{2m}|\lambda_{a}|^{2r}\leq \Tr C^r(C^r)^T,
\end{equation}
where $\lambda_{a}$ denotes one eigenvalue, and there are $2m$ such eigenvalues.
Note that $\brb{C^r(C^r)^T}_{i\to j,i\to j}$ is contributed by paths connecting edge $i\to j$
with distance $r$-step away:
$$
\brb{C^r(C^r)^T}_{i\to j,i\to j}=\sum_{U}\prod_{(x\to y) \in U}\tanh^2(\beta J_{x\to j}),
$$
where $U$ denotes one such path and $(x\to y)\in U$ denotes edges belonging to path $U$.
When $n$ is large, and couplings $\bre{J_{ij}}$ are i.i.d. distributed, for each path $U$ we have
$$
\prod_{(x\to y) \in U}\tanh^2(\beta J_{x\to j})=\brc{\tanh^2(\beta J_{i\to j})}^r,
$$
where average $\brc{\cdot}$ is taken over realizations of couplings, 
and there are $\hat c^r$ such paths with $\hat c$ denoting excess degree 
$$\hat c=\sum_{k}\frac{p(k)k(k-1)}{\sum_kp(k)k}=\frac{<k^2>}{c}-1.$$
By taking expectation we have
$$\mathbb{E} \Tr C^r(C^r)^T=2m\hat {c}^r\brc{\tanh^2(\beta J_{i\to j})}^r.$$ 
Then relation \eqref{eq:ineq} can be written as
\begin{equation}
	\mathbb{E}\bra{|\lambda|^{2r}}\leq \bra{\hat c\brc{\tanh^2(\beta J_{i\to j})}}^r,
\end{equation}
and it holds for any $r$. Last equation tells us that almost all $C$'s eigenvalues are confined by a disk of 
radius $R$ in the complex plan, with
\begin{equation}\label{eq:R}
	R=\sqrt{\hat c\brc{\tanh^2(\beta J_{ij})}}.
\end{equation}
The disk containing almost all the eigenvalues means the density of eigenvalues 
$$\rho(\lambda)=\frac{1}{2m}\sum_{i\to j}\delta(\lambda-\lambda_{i\to j})$$
is non-zero inside the disk, and is zero outside the bulk. Note that 
there could be outlier eigenvalues outside the bulk with zero density.

There is another way to understand the edge of bulk, in the sense
of noise propagation on trees: consider some 
random noise with mean $0$ and unit variance are put on the leaves of a tree; by iterating $C$, the noise 
may be propagated to the root of the tree with non-vanishing variance, if 
$\hat c\brc{\tanh^2(\beta J_{i\to j})}>=1.$ Thus radius of disk $R>1$ tells us that random noise could 
make Edwards-Anderson order parameter \cite{Edwards1975} of the system 
finite, which is a sign showing that system is in the ``spin glass'' state, and replica symmetry is 
breaking. So $R$ tell us where is the spin glass state and its relative strength to other possible states.
$R=1$ point is also known as the de Almeida-Thouless local stability condition~\cite{Almeida1978}, 
the Kesten-Stigum bound~\cite{kesten1966,kesten1966a}, 
or the threshold for census or robust reconstruction~\cite{Mezard2009,janson2004robust}.

If there are real-eigenvalues outside the bulk, they may tell us other possible state of the system, which 
correspond to other BP solutions.
For example, if all the couplings are positive, it is easy to see that 
there could be one positive eigenvalue associated with an eigenvector with most of its 
entries positive. We can approximately recover this eigenvector by the following scheme: assume the 
graph is a tree with all its nodes associated with spin $S_k=1$, or equivalently on the edge 
associated with leaves of the tree $S_{k\to l}=1$. 

By $r$ steps iterating of $C$, 
we define an ferromagnetic ``approximate'' eigenvector  $g^{(r)}$ as
\begin{equation}
	g^{(r)}=\frac{1}{\mu^r}C^rS.
\end{equation}
If with $r\to \infty$ the vector $(g^{(r)})_{i\to j}$ is still correlated with $S_{k\to l}$, the 
all-one vector on the leave, system has a ferromagnetic state where information is preserved during iterating.
On a tree, last equation can be written as contribution of paths $U$ connecting edge $i\to j$ and edge 
$k\to l$ distance $r$ away:
\begin{equation}\label{eq:g}
	(g^{(r)})_{i\to j}
	= \frac{1}{\mu^r}\sum_{U}\prod_{(x\to y) \in U} \tanh(\beta J_{xy}),
\end{equation}
where $(x\to y) \in U$ denotes edges belong to path $U$, and there are $\hat c^r$ such paths.

When $n$ and $r$ are large, if we assume the self-average property,
\begin{equation} 
	\prod_{\brc{x,y}}\tanh(\beta J_{xy})=\brc{\tanh(\beta J_{xy})}^r,
\end{equation}
Eq.~\eqref{eq:g} can be written as
\begin{equation}
	(g_{i\to j})^{(r)}
	= \frac{1}{\mu^r}\sum_{U}\brc{\tanh(\beta J_{xy})}^r,
\end{equation}
where $\brc{\cdot}$ is again taking average over realizations of couplings.

Iterating $C$ once gives
\begin{equation}
	\bra{Cg^{(r)}}_{i\to j}=\frac{1}{\mu^r}c^{r+1}\brc{\tanh(\beta J_{xy})}^{r+1}
	=\mu g_{i\to j}^{(r)},
\end{equation}
and the eigenvalue is
\begin{equation}\label{eq:mu}
	\mu=\hat c\brc{\tanh(\beta J_{ij})}.
\end{equation}
$g_{i\to j}$ is indeed an eigenvector of $C$ if $g_{i\to j}^r-g_{i\to j}^{r+1}$ goes to zero when $r$ 
is large. In another words, $g_{i\to j}$ is an eigenvector if the noise propagation 
is slower than propagation of information $\bre{1,1,..,1}$, that is
$ \mu > R$, which asks the informative eigenvalue of ferromagnetic state out of the bulk.
Besides the ferromagnetic eigenvalue, there could be other real-eigenvalues representing other states of
the system.  So the competition of real-eigenvalues and $R$ gives us the information of the phases and 
phase transitions of system.

We note that $R$ and $\mu$ have been already computed in \cite{Mooij05}, using a different technique. However 
authors in \cite{Mooij05} did not claim or show the sharp edge of bulk of spectrum. Another reason 
that we keep the detailed computation of $\mu$ and $R$ in this paper
is that we need the technique in Sec.~\ref{sec:hop} to compute the spectrum of non-backtracking 
matrix of Hopfield neural networks.

\section{Examples on ferromagnet and spin glasses}\label{sec:examples}
In this section we apply the 
non-backtracking operator to Ising model with specific distributions of couplings. One aim is to see 
whether non-backtracking operator recovers the known results of phase transitions on those systems.
\subsection{Ferromagnetic Ising model on random graphs}
For first example we consider ferromagnetic Ising model, which has $J_{ij}=1$.
If we consider ferromagnet on Erd\H{o}s-R\'enyi random graphs where $\hat c=c$, result given by 
non-backtracking matrix would be 
quite accurate, due to the locally tree-like structure of the topology.
From the analysis of last section, for this model we have 
$$R_{F}=\sqrt{c}\tanh\beta$$ and $$\mu_{F}=c\tanh\beta,$$ 
thus $\mu_F$ is always larger than $R_F$ with finite $\beta$ and $\hat c>1$.
It means this system can never be in spin glass state, it has to be in either ferromagnetic state 
or paramagnetic.
And at $c\tanh\beta>1$, which implies that at 
\begin{equation}
	\beta=\tanh^{-1}\frac{1}{c},
\end{equation}
system undergoes a transition from paramagnetic to ferromagnetic phase.

\begin{figure}
   \centering
    \includegraphics[width=0.45\columnwidth]{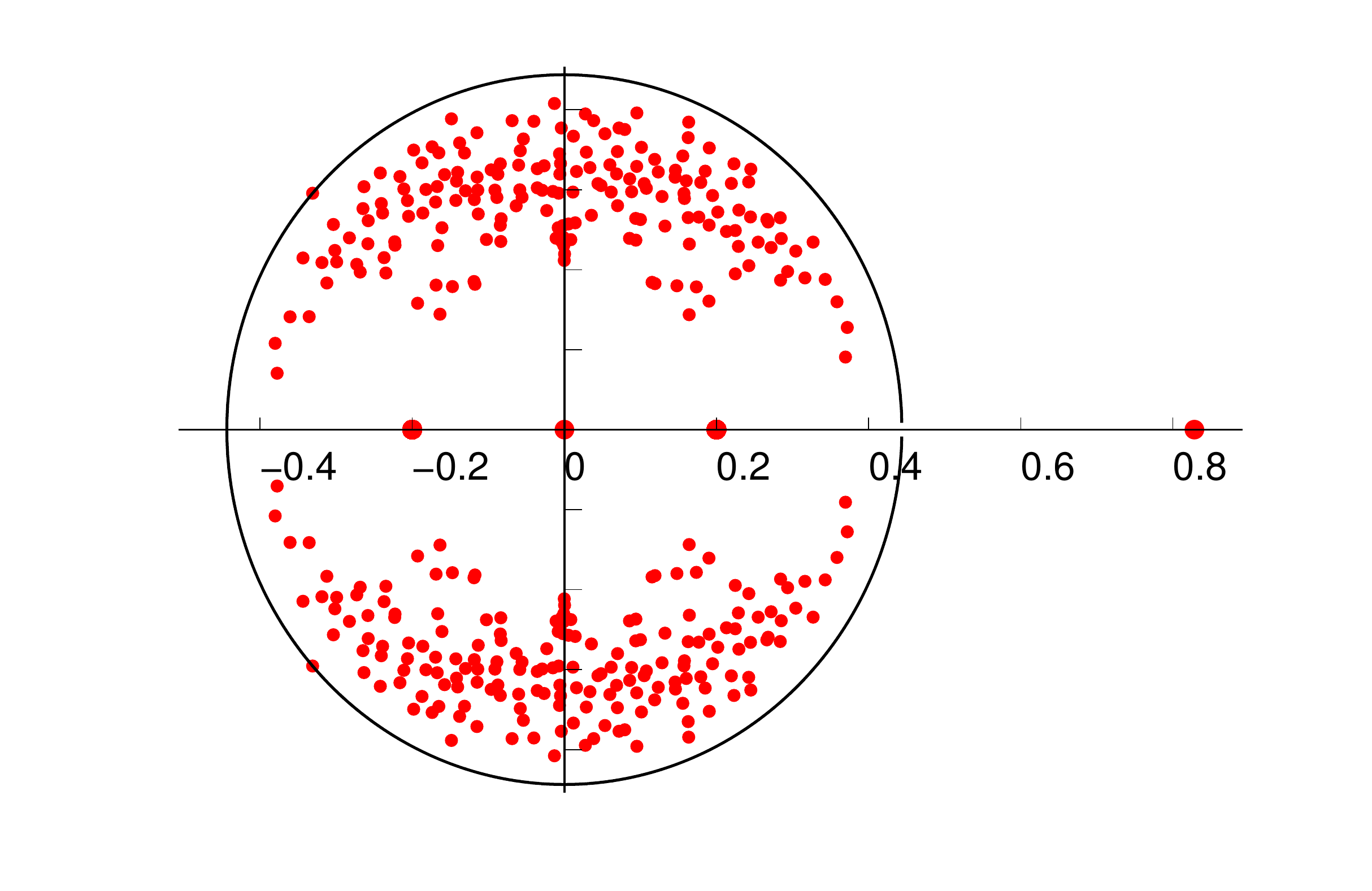} 
	\includegraphics[width=0.45\columnwidth]{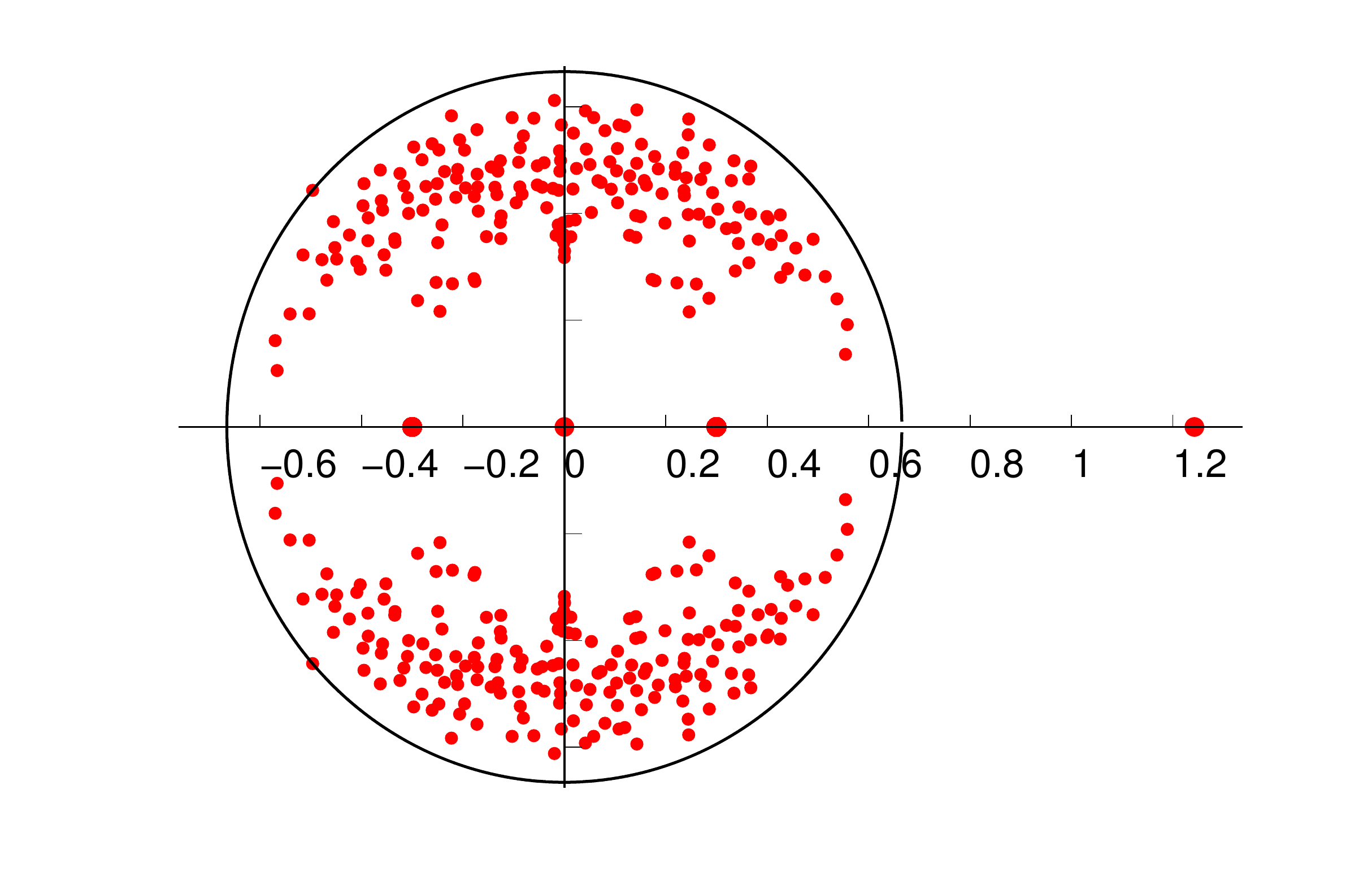}
	\caption{	
	Ising model on a random graph with $n=200, c=4$ and $\tanh(\beta)=0.2 < 1/c$ (left) 
	and $\tanh(\beta)=0.3 > 1/c$ (right)	\label{fig:ising}}
\end{figure}

Note that $C$ can be expressed as
\begin{equation} \label{eq:ferro}
	C=B\tanh\beta,
\end{equation} where $B$ is the non-backtracking matrix 
on the graph. From \cite{Krzakala2013} we know that $B$ has the largest eigenvalue $\lambda_1^B=\hat c$, so as a 
sanity check, one can verify that $\mu=\tanh(\beta)*\lambda_1^B$. 
In Fig.~\ref{fig:ising} spectrum of $C$ is plotted in the complex plane for Ising model on a random graph 
with two $\beta$ values, one is in paramagnetic state and the other one is in ferromagnetic state.
As shown in the figure, with $\tanh(\beta)<\frac{1}{c}$, the largest eigenvalue of $C$ is smaller than 
one, which tells us that the paramagnetic fixed-point is stable. With $\tanh(\beta)>1/c$, the largest 
eigenvalue of $C$ is greater than $1$, telling us the paramagnetic fixed-point is unstable towards the 
direction of ferromagnetic state.
It has been shown in \cite{Krzakala2013} that non-trivial part of $B$'s spectrum is can be obtained from 
a $2n\times 2n$ matrix
$$B'=
 \left( \begin{array}{cc}
	0 & D-I\\
	-I & A 
\end{array} \right),$$
where $D$ is diagonal matrix with degree of nodes on the diagonal entries, $I$ is the identity matrix and
$A$ is the adjacency matrix.
Thus non-trivial part of $C$'s spectrum for ferromagnet can also be computed with less effort.

\subsection{Ferromagnetic Ising model on networks with a community structure}
If the network has a community structure, e.g. the number of edges connecting same group
is much larger than number of edges connecting to different groups,
besides the eigenvector pointing in direction to all-one vector, there could be other eigenvectors
with some positive and some negative entries, with signs representing group memberships.

Here we consider networks that generated by a generative model named Stochastic Block Model (SBM) or planted
partition model \cite{Decelle2011,Decelle2011a}.
SBM is one of the widely used model to generate benchmark networks containing 
community structure. In the model, there are $q$ groups of nodes, each node $i$ has a pre-assigned 
group label $t^*_i$, edges are generated independently according to a $q\times q$ matrix $p$ 
by connecting each pair of nodes $\brc{ij}$ with probability $p_{t^*_i,t^*_j}$. Here 
for simplicity we discuss the commonly studied case that matrix $p$ has only two distinct 
entries, $p_{\text{ab}}=p_{\text{in}}$ if $a=b$ and $p_{\text{out}}$ if $a\neq b$, and we use 
$\epsilon=p_{\text{out}}/p_{\text{in}}$
to denote the ratio between these two entries. A phase transition at $\epsilon^*$
\cite{Decelle2011,Decelle2011a}
has been discovered in this model that finite amount of information about 
planted partition $\{t^*\}$ can be inferred with $\epsilon<\epsilon^*$, and one can 
do the optimal inference of SBM to recover information of $\{t^*\}$ all the way down to the 
transition. So it is a good benchmark for testing performance of community detection algorithms.
Here we consider the simple two-group case with equal group sizes. 
From \cite{Krzakala2013} we know that in the detectable regime, edge of bulk of matrix $B$ is $\sqrt c$, 
it has two real eigenvalues outside the bulk, the first one $\lambda_1=1$ corresponds to the eigenvector 
with entries having same sign. And the second one $\lambda_2=\frac{1-\epsilon}{1+\epsilon}c$ corresponds 
to the eigenvector correlated with community structure. According to Eq.~\eqref{eq:ferro}, edge of 
bulk of matrix $C$ on the network is $$R=\sqrt c\tanh\beta,$$ first eigenvalue is $\mu_1=c\tanh\beta$,
and the eigenvalue corresponding to the community structure is 
$$\mu_2=\frac{1-\epsilon}{1+\epsilon}c\tanh\beta.$$
So if $\mu_2>1$, the community structure in the network is detectable 
by Ising model (by excluding the ferromagnetic state in some way). 

Recall that Belief Propagation we used to derive the non-backtracking matrix at the beginning of the paper 
is nothing but an iterative method to 
decrease Bethe free energy; and a fixed point of Belief Propagation is a local minimum of Bethe free energy
 \cite{Yedidia2001}. Obviously paramagnetic solution is always one local minimum of Bethe free energy.
As we discussed above, eigenvalues larger than one in non-backtracking matrix tells us 
the instability of this paramagnetic fixed-point, which is also the 
instability of this local minimum of Bethe free energy.
There is another way to study the instability of a minimum of free energy, which is 
Hessian matrix of Bethe free energy (we call it Hessian for simplicity) evaluated at paramagnetic fixed-point 
\cite{saade2014a,Ricci-Tersenghi2012}, 
which reads
\begin{equation}
	H_{ij}=\left. \frac{\partial ^2 F_{\text{Bethe}}}{\partial \psi_i\partial \psi_j}
	\right |_{\psi=\frac{1}{2}}=\brb{ 1+\sum_k\frac{\tanh^2(\beta J_{ik})}{1-\tanh^2(\beta J_{ik})}}\delta_{ij}
-\frac{\tanh(\beta J_{ij})}{1-\tanh^2(\beta J_{ij})}.
\end{equation}
Detailed derivations can be found at e.g. \cite{Ricci-Tersenghi2012}. For ferromagnets
last equation reduces to
\begin{equation}
	H=I+\frac{\tanh^2\beta}{1-\tanh^2\beta}D-\frac{\tanh\beta}{1-\tanh^2\beta}A.
\end{equation}
The paramagnetic state is stable if this matrix is positive semidefinite. If it
has negative eigenvalues, there must be other minimums of free energy at some point 
in the phase space, in direction that pointed by components of eigenvectors 
corresponding to negative eigenvalues. This scheme has been used in 
\cite{Krzakala2013}
 to do spectral clustering on graphs where the Hessian is expressed as a function of $r$, 
with $r=\frac{1}{\tanh\beta}$ in our settings. 

Given $\beta$ and the graph, we have the following picture on the non-backtracking matrix and 
Hessian matrix:
if $\mu_1<1$, only paramagnetic fixed-point is stable, Hessian is positive 
semi-definite. If $\mu_1>1$ and $\mu_2<1$, paramagnetic fixed-point is unstable towards the ferromagnetic 
state, Hessian has one negative eigenvalues. If $\mu_2>1$, paramagnetic fixed-point is unstable 
towards both the ferromagnetic state and the planted configuration (the community structure), Hessian 
has two negative eigenvalues. Then we see that non-backtracking matrix works equivalently to Hessian 
in determine the phases of ferromagnets. But we note that computing spectrum of Hessian is much faster than 
non-backtracking matrix, because Hessian is symmetric and size of Hessian is smaller ($n\times n$ instead 
of $2m\times 2m$ in non-backtracking matrix).

If we are not interested in the phase of Ising model but only the problem of node clustering 
in a given network, we can choose $\beta$ as we want. As we shown above, $\beta$ does not influence 
the shape of spectrum of non-backtracking matrix, if $\mu_2>R$, the second eigenvalue is always outside 
the bulk, thus we can always check whether there is a second eigenvalue outside the bulk to determine 
number of groups and use the eigenvector associated with it to detect the community structure.
This is the idea explored in \cite{Krzakala2013} which uses matrix $B$ in doing spectral clustering.

The situation is different for Hessian, if $\beta$ is too small, Hessian is positive semidefinite, it looses
its ability of determining number of groups in network. So it is important to select a good $\beta$ value 
that gives negative eigenvalues.
The authors of \cite{saade2014a} suggested to chose $r=\sqrt{c}$, equivalent to
set $\tanh\beta=\frac{1}{\sqrt c}$ in our settings. As we discussed above, $r=\sqrt{c}$ is the 
point that edge of bulk reaches $1$, thus if there is a second real-eigenvalue of $C$ 
outside the bulk and is greater
than $1$, or equivalently a stable state correlated with community structure in Ising model, one should be 
able to find a second negative eigenvalue of $H$ with $r=\sqrt{c}$.

On the one hand, detecting number of groups using real eigenvalues of $B$ outside the bulk is easier than 
tunning $\beta$ value to look for negative eigenvalues of $H$, especially on real-world networks.
But on the other hand, tunning $\beta$ gives $H$ ability to optimize accuracy of detected communities, at 
least in synthetic networks.

\subsection{SK model}
Spin glass models were proposed to explain magnetic systems with frozen structural disorder, by 
introducing ``frustration'' using a particular distribution of couplings. Now theory of 
spin glasses e.g. replica symmetry breaking and cavity method find many application is many fields.
In this section we consider non-backtracking matrix of two spin glass models, namely Sherrington-Kirkpatrick
model \cite{Sherrington1975} and Viana-Bray model \cite{Viana1985}. It is well known that those spin 
glass models have paramagnetic phase at high temperature and spin-glass phase at low-temperature where 
both magnetization and Edwards-Anderson parameter are non-zero, and the replica symmetry is
broken \cite{Mezard1987}. 
First we consider Sherrington Kirkpatrick model which is defined on 
fully connected graph with couplings following Gaussian distribution:
$$P(J_{ij})=\frac{1}{\sqrt{n}}\mathcal{N}(0,1).$$
From Eq.~\eqref{eq:R} and Eq.~\eqref{eq:mu}, we have $$\mu_{\text{SK}}=0$$ and
\begin{equation} \label{eq:RSK}
	R_{\text{SK}}=\sqrt{n\brc{\tanh^2(\beta J_{ij})}}\approx \sqrt{n\brc{\beta^2 J_{ij}^2}}=\beta.
\end{equation}
So the paramagnetic solution is stable as long as $\beta\le 1$. And system enters spin glass state with 
$\beta>1$. These are the well known results.

\begin{figure}
   \centering
    \includegraphics[width=0.6\columnwidth]{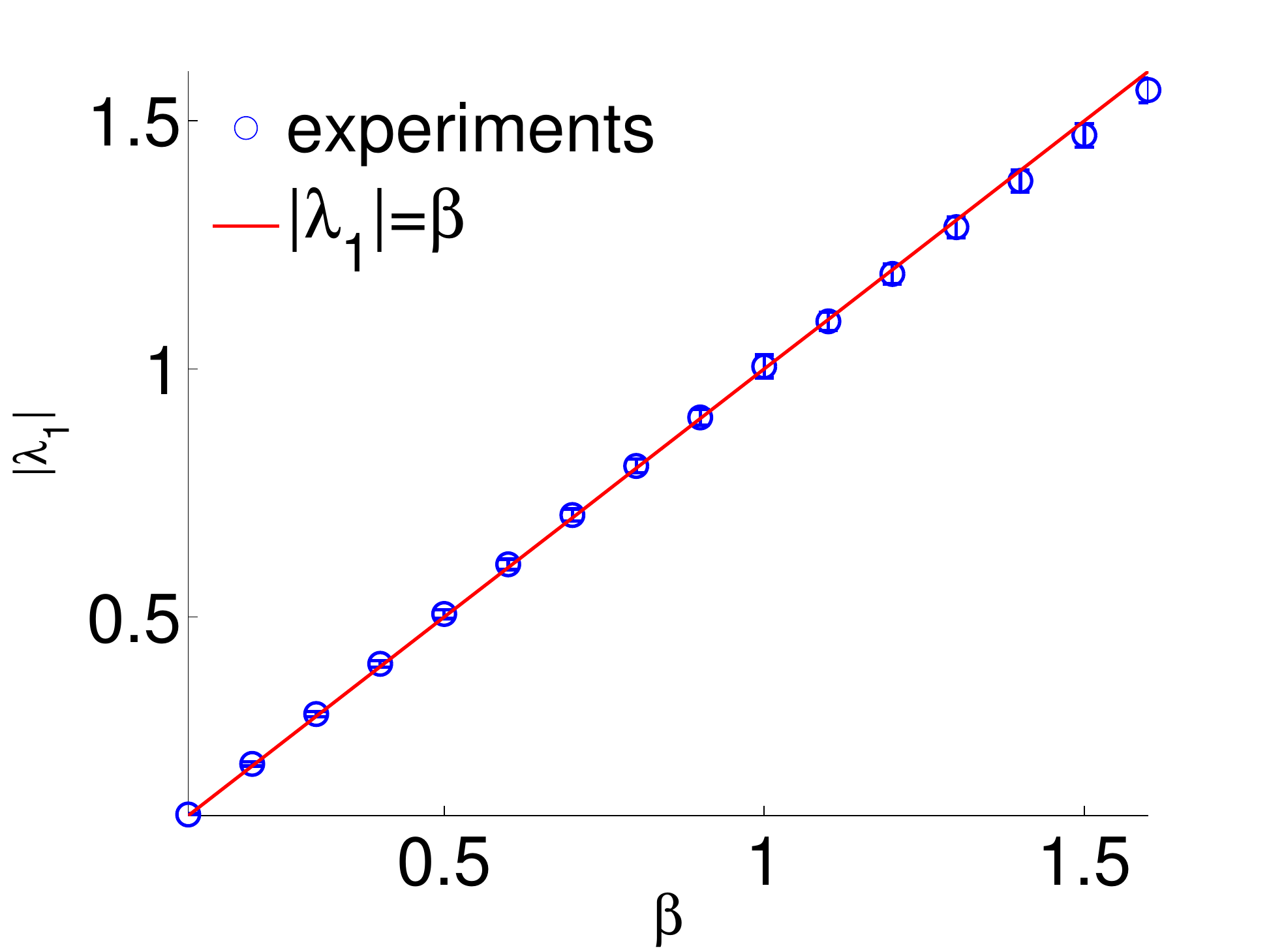} 
	\caption{	
	Absolute value of the largest eigenvalue of non-backtracking operator, $|\lambda_1|$, as a 
	function of $\beta$ for SK model with $n=60$. Each point is averaged over $10$ instances.
		\label{fig:sk_beta}}
\end{figure}

To test theory (Eq.~\eqref{eq:RSK}), in Fig.~\ref{fig:sk_beta} we plot the edge of bulk 
(absolute value of the largest eigenvalues) as a function of $\beta$ for networks with $n=60$. We can 
see that it consistent well with $|\lambda_1|=\beta$. 

\begin{figure}
   \centering
    \includegraphics[width=0.35\columnwidth]{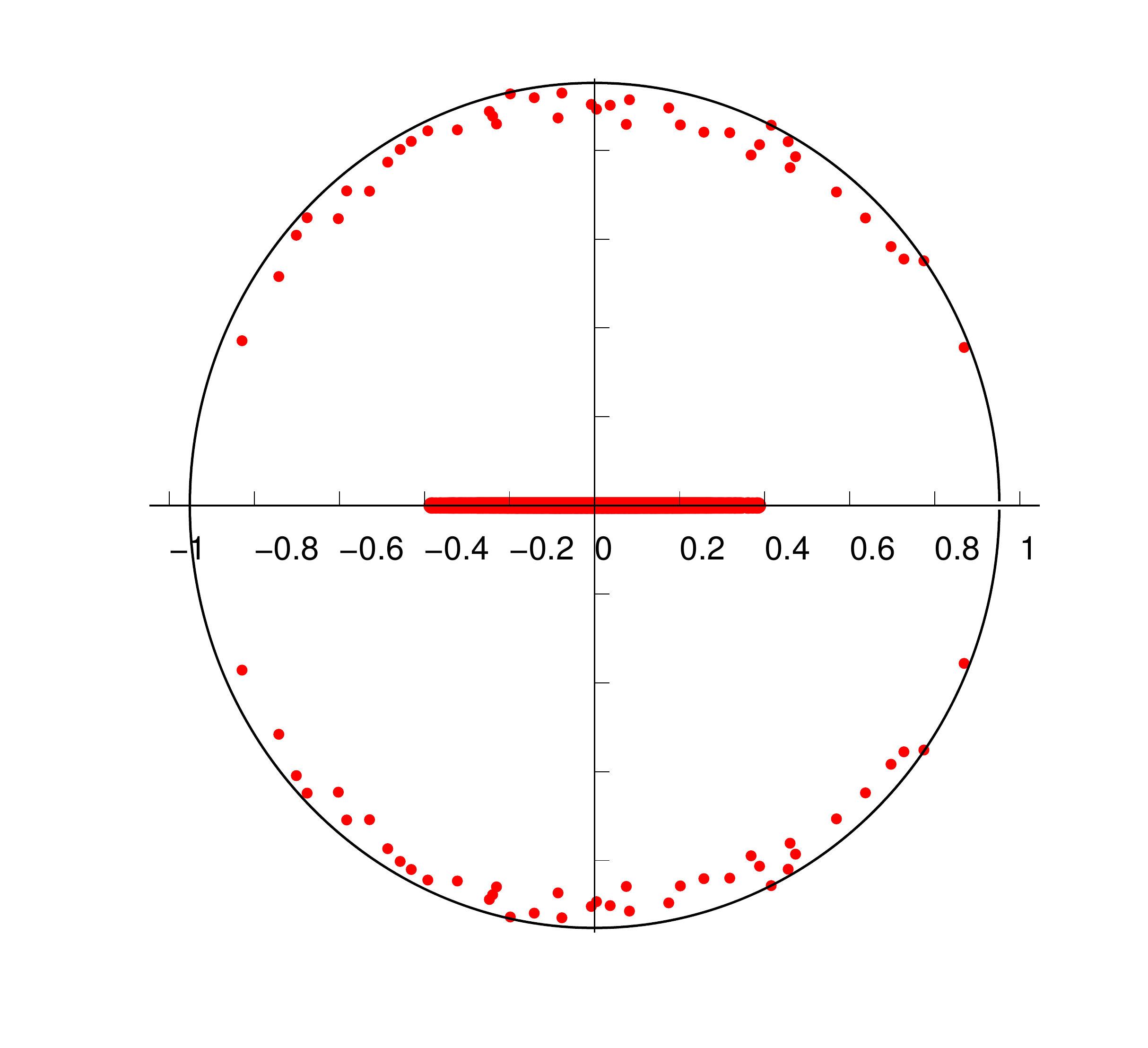} 
	\includegraphics[width=0.35\columnwidth]{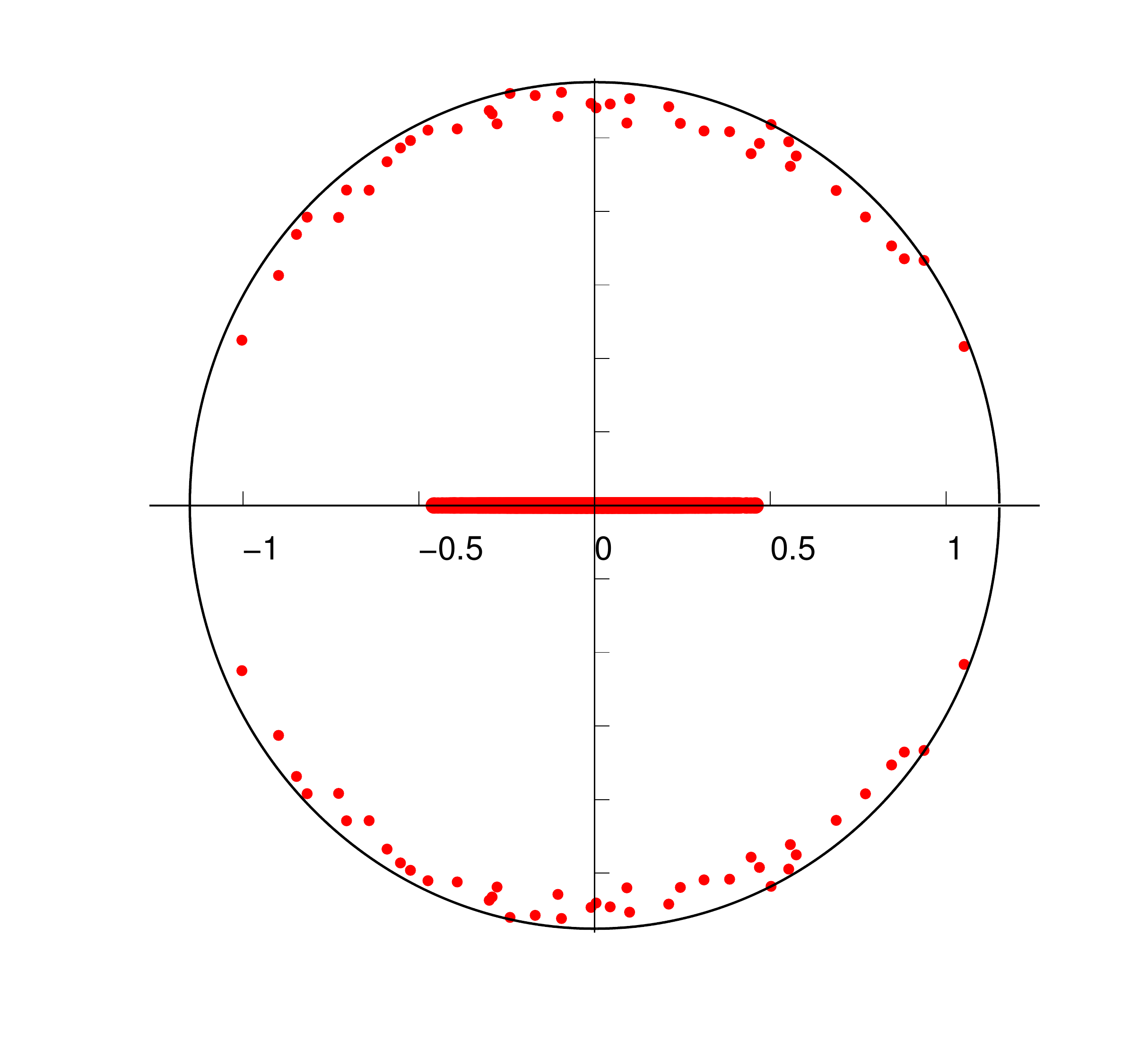}
	\caption{	
	Spectrum (in the complex plane) of non-backtracking matrix of 
	SK model with $n=40$ and $P_{J_{ij}}=\frac{1}{\sqrt{n}}\mathcal{N}(0,1)$.
	In left panel $\beta=0.9$, system is in paramagnetic phase and in right panel
	$\beta=1.1$, system is in spin glass phase.\label{fig:sk}}
\end{figure}

The spectrum of $C$ on SK model with two $\beta$ values are also shown in Fig.~\ref{fig:sk}.
We can see that with $\beta<1$, the edge of bulk is smaller than $1$, which means system is in paramagnetic
phase.
In the right ,with $\beta>1$, the edge of bulk is greater than $1$, system is in the spin glass phase.
Moreover, from Eq. \eqref{eq:mu} we can also see that for the fully connected model 
distribution of couplings are not important, what really matters to the spin glass transition 
is the variance of couplings.

From the last section, we see that Hessian matrix of Bethe free energy works as well as 
non-backtracking matrix in
detecting paramagnetic to ferromagnetic transition in ferromagnets, but it does not work well in 
SK model. Our numerical result (as well as experimental results in \cite{Mooij05}) show that Hessian 
matrix for SK model is positive definite, thus it does not tell us where system encounters the spin 
glass transition.

\subsection{Viana-Bray model}
\begin{figure}
   \centering
   \includegraphics[width=0.4\columnwidth]{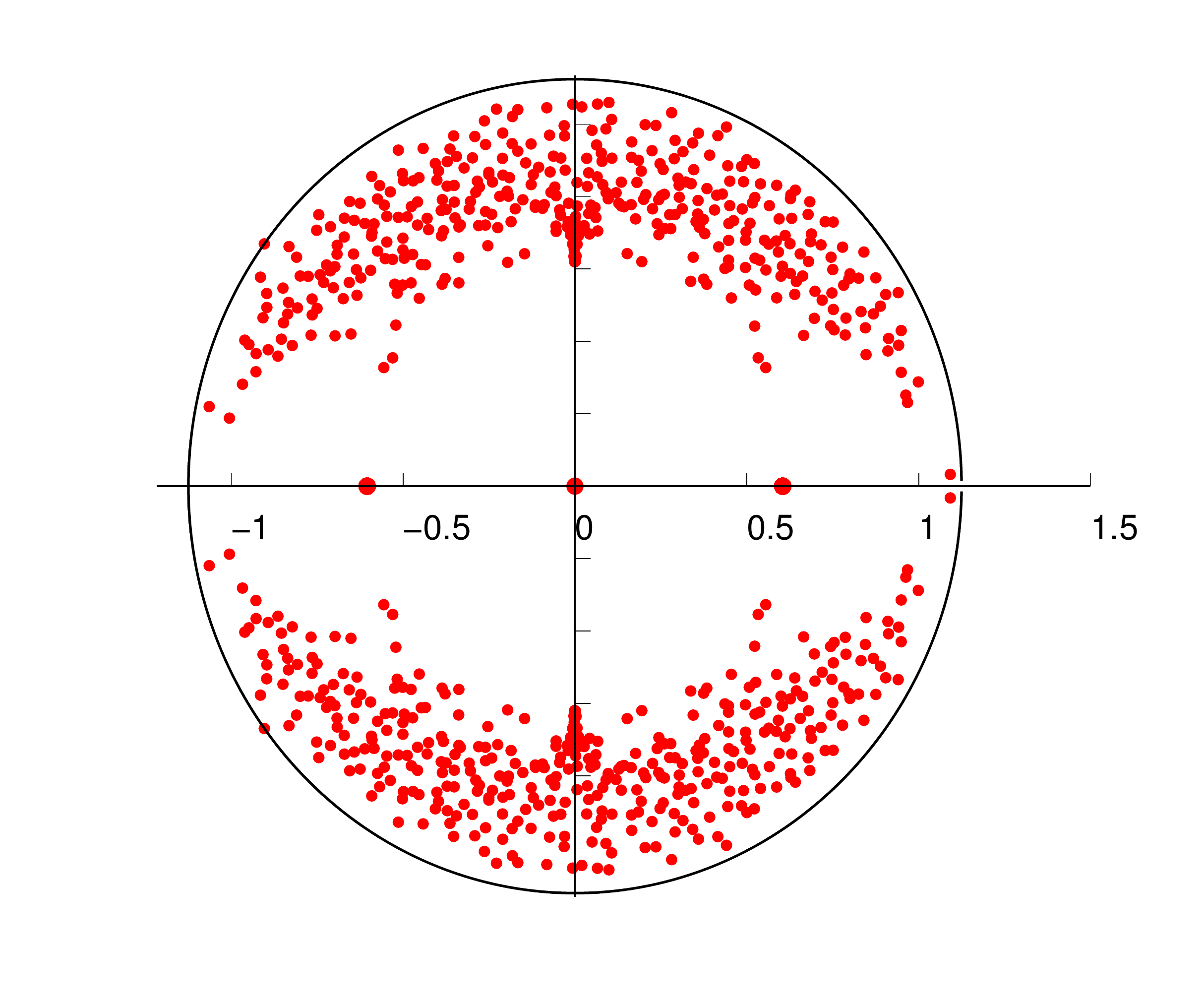}
   \includegraphics[width=0.4\columnwidth]{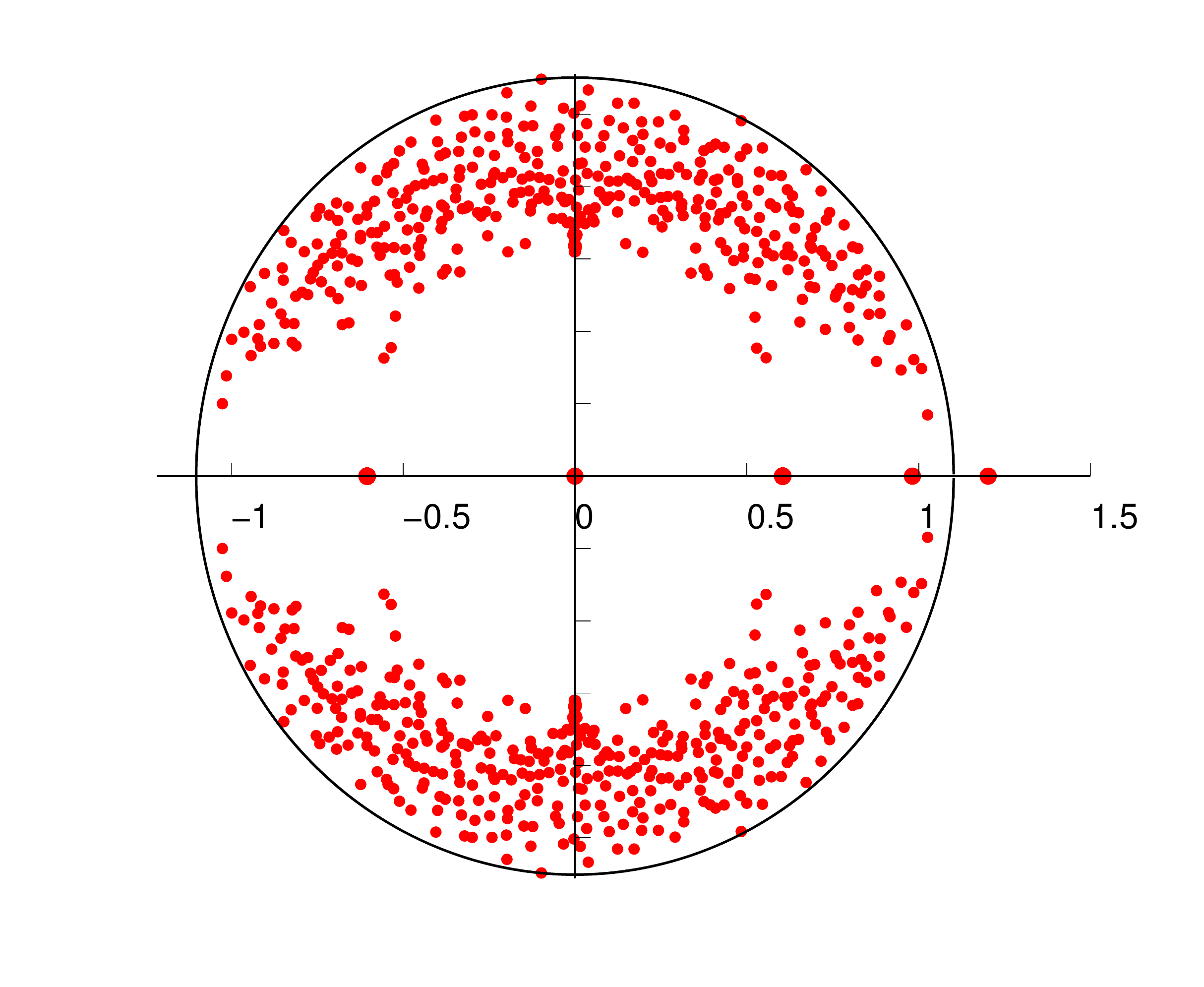} 
	\caption{	
	Spectrum (in the complex plane) of non-backtracking matrix of Viana-Bray model on a 
	graph with $n=500$, $c=3$. In the left panel 
	$\beta=0.7>\tanh^{-1}\frac{1}{\sqrt c}$, $p^+=0.5$, system is in spin glass phase and in the 
	right panel $p^+=0.7$, system is in ferromagnetic phase	\label{fig:vb}}
\end{figure}

Viana-Bray model \cite{Viana1985}, also called $\pm J$ model, 
is a spin glass model defined on random graphs. It is similar to 
the ferromagnets on random graphs but with couplings taking $+1$ and $-1$ randomly which give frustration
to the system.
$$p(J_{ij})=p^+\delta_{J_{ij}}+(1-p^+)(1-\delta_{j_{ij}})$$
By varying $p^+$ and $\beta$, the system can be in paramagnetic phase, spin glass phase or 
ferromagnetic phase.
From ~Eq.\eqref{eq:R}\eqref{eq:mu} we have
$$\mu=2p^+\tanh\beta-1,$$ and $$R=\sqrt{c}\tanh\beta.$$ 
Obviously system undergoes paramagnetic to 
spin glass transition at $\beta=\tanh^{-1}\frac{1}{\sqrt{c}}$ independent of $p^+$.
And may undergo paramagnetic to ferromagnetic
transition if $\mu>R$. So our result recovers the phase boundaries derived using replica methods 
in \cite{Viana1985}.
In Fig.~\ref{fig:vb} we plot the spectrum of non-backtracking matrix for Viana-Bray model in the complex plane
for system in two phases, one in spin glass phase where there is no real eigenvalues outside the bulk,
and the other one in ferromagnetic phase where there is one real eigenvalue outside the bulk and is 
greater than $1$.

\section{Non-backtracking operator for Hopfield model and controlling of neural networks}\label{sec:hop}
Hopfield model is a classic model for associative memory, it stores patterns by couplings of Ising model.
We assume $P$ random binary patterns 
$\xi_i^\mu=[-1,+1]$ with $i=[1,...,n]$, 
$\mu=[1,...,P]$ stored in network by Hebb's rule \cite{hebb1949}:
$$J_{ij}=\frac{1}{P}\sum_{\mu=1}^P\xi_i^\mu\xi_j^\mu.$$
If $P$ patterns are memorized successfully, they can be retrieved by Glauber dynamics \cite{Glauber1963}: at time $t$ one neuron $i$ is randomly
selected and its value $\sigma_i^t$ is updated according to the probabilistic rule
$$p(\sigma_i^t=\sigma)=\frac{e^{\beta\sum_{j\in\partial i}J_{ij}\sigma\sigma_j^{t-1}}}
{2\cosh{\beta\sum_{j\in\partial i}J_{ij}\sigma_j^{t-1}}}.$$
From a random configuration, Glauber dynamics could converge to set of configurations
that correlated with one pattern.
If all patterns are memorized successfully in the network, each pattern will behave like an attractor 
to the Glauber dynamics, with different basin of attractions.
That is why Hopfield model is categorized into attractor neural networks \cite{amit1992}. The reason that
patterns behave as attractors to the dynamics is that in Hebb's rule, 
each pattern corresponds to a minima of free energy,
and Glauber dynamics, which a ``spin-flip'' algorithm that converges to equilibrium, will 
trap into one of those local minimum of free energy.

The original Hopfield network was defined on fully-connected network, and its statistical 
mechanics has been studied \cite{Amit1985a,Amit1985} using replica method. 
However fully-connected model is not biologically realistic. In this paper we will focus on 
its more biologically-realistic version, which is Hopfield model on a random graph. 
The model is also called finite-connectivity Hopfield model and its statistical mechanics
has also studied using replicas \cite{Wemmenhove2003}, and phase boundaries were analyzed. 
The phase diagram of Hopfield model (both fully-connected and on a random graph) is composed of 
three parts: at high temperature every neuron behaves the same, magnetization is zero, 
system is in the paramagnetic phase; 
at a low temperature and with a low number of patterns, patterns are attractive to Glauber dynamics and 
system is in ``retrieval'' or ``memory'' phase; at a low temperature and a high number 
of patterns, network is confused by so many patterns thus none of the pattern is memorized successfully, Glauber 
dynamics will go away even started from one of the patterns, system is in spin glass phase. 

Note that replica results were derived for the model at thermodynamic limit averaged over disorder
(realizations of patterns), thus it is hard to apply the replica result to a single instance of network,
to do tasks such as retrieving patterns and determining number of patterns. 
The classic way to do such tasks is Glauber dynamics. However on large networks it is time consuming,
and the result depends on the initial configuration of the dynamics. And it is hard
for Glauber dynamics to determine how many patterns stored in the network, because one can not 
explore all the initial configurations. 

A faster way to retrieval patterns is running BP on the instance. But there is still a
drawback that it is difficult to answer how many patterns there are in the network, 
since explore all possible initial messages for BP is hard. 

In the rest part of the section we will study how to do the pattern retrieval task using non-backtracking
matrix. It recovers replica result at thermodynamics limit easily, and on single instances it has advantages 
over Glauber dynamics and BP for speed as well as for determining number of patterns.

First we study the spectrum of non-backtracking operator at thermodynamic limit.
From Eq.~\eqref{eq:R}, the edge of bulk of Hopfield model, 
which tells us where is the spin glass phase, is expressed as 
\begin{equation}\label{eq:hop:R} 
	R_{\text{H}}=\sqrt{\hat c\brc{\tanh^2\bra{\frac{\beta}{P} \sum_{\mu=1}^P\xi^\mu}}},
\end{equation}
the averaging is taken over random patterns.

Obviously there is no stable ``ferromagnetic'' state since if we iterate $C$ 
with all-one configuration $\{1,1,...,1\}$ on leaves of the tree,
we get 
$$\hat c\brc{\tanh\bra{\frac{\beta}{P} \sum_{\mu=1}^P\xi^\mu}}=0.$$ 
However, note that if instead of all-one configuration, we put a
pattern in the leaves of the tree and do iterating, the information of the pattern could be preserved during
iterating. It is very similar to initialize belief propagation by messages correlated with one 
pattern, or running Glauber dynamics from a pattern. 
Analysing iterating with a random pattern on leaves is difficult, 
but we can show it in an equivalent but easier way, by focusing on 
one of the stored patterns, let us say $\xi^1$ without loss of generality. 
Since patterns are chosen randomly and independently, we can do 
a gauge transformation to all the patterns which transforms $\xi^1$ to all-one configuration, 
and transforms other patterns to $$\hat \xi_i^{\mu}=\xi_i^1\xi_i^\mu.$$ 
It is easy to 
see that this transformation does not change distance between patterns, thus does not change the property of 
the system. Original patterns can be recovered easily by product $\xi_i^1$ again:
$$\xi_i^{\mu}=\xi_i^1\hat \xi_i^\mu.$$ 
Under this transformation, couplings become 
$$\hat J_{ij}=\frac{1}{P}\brb{
1+\sum_{\mu=2}^P\hat\xi_i^\mu\hat \xi_j^\mu}.$$
Obviously now couplings are biased to positive, and there could be a ferromagnetic state in the 
transformed system. Then we can treat $\frac{1}{p}$ term as a ``signal'' term 
that tells us information about all-one configuration, and term 
$\sum_{\mu=2}^P\hat\xi_i^\mu\hat \xi_j^\mu$ as a ``noisy'' term that gives some cross-talk noise to signal 
term. If there is only one pattern, cross-talk noise is $0$, the problem essentially goes back 
to the ferromagnets. When number of pattern is small, noisy term has small fluctuation thus signal 
remains clear during iterating. 

It is easy to see that the eigenvector correlated to the all-one vector in the 
transformed system is 
related to the eigenvector correlated with the first pattern in the original system.
And the eigenvalue associated with the eigenvector correlated with all-one
vector in the transformed system is the same with the eigenvalue associated with 
eigenvector correlated with first pattern in original system.
With $n\to \infty$, this eigenvalue can be written as 
\begin{equation}
	\mu_{\text{H}}=\hat c\brc{ \tanh\frac{1}{P}\bra{1+\sum_{\mu=2}^P\hat\xi^\mu}}.
\end{equation}
So setting $\mu_{\text{H}}=1$ gives paramagnetic to retrieval transition and $R_{\text{H}}=1$ 
gives paramagnetic to spin glass transitions. 

If we write last equation in the form of 
\begin{eqnarray}
	\mu_{\text{H}}&=&\hat c\brc{ \tanh\frac{1}{P}\bra{\xi^1\xi^1+\sum_{\mu=2}^P\hat\xi^\mu}}\nonumber\\
	&=&\hat c\brc{ \xi^1\tanh\frac{1}{P}\bra{\xi^1+\sum_{\mu=2}^P\hat\xi^\mu}},
\end{eqnarray}
we can see that the expression for phase boundaries are in agreement 
with result obtained by replica method \cite{Wemmenhove2003}, considering
a little different definition of the model. 
Moreover, with patterns chosen randomly at thermodynamic limit, we have
\begin{eqnarray}
	\mu_{\text{H}}&=&\frac{\hat c}{2^{P}}\sum_{s=0}^{P-1}{{P-1}\choose{s}}\tanh\bra{\beta\frac{P-2s}{P-1}}\nonumber\\
	R_{\text{H}}&=&\sqrt{ \frac{\hat c}{2^P}\sum_{s=0}^P{{P}\choose{s}}\tanh^2\bra{\beta\frac{P-2s}{P}}}.
\end{eqnarray}

If we do the above gauge transform on every pattern, we would have $P$ such eigenvalues.
But note that though we studied the $\mu_{\text{H}}$ and $R_{\text{H}}$ with the gauge transform,
it is only a trick to study analytically in an easier way the eigenvalues. In a single instance,
we do not need to do such transform, there should be $P$ eigenvalues corresponds to $P$ patterns.
And on the given network, once we find real eigenvalues outside the bulk, we can use eigenvectors
associated with those eigenvalues to retrieve patterns. 

For more detail, after we have an eigenvector $v_{j\to i}$, first we obtain 
the vector $S$ containing summation of incoming values from edges for nodes, from an eigenvector $v$,
$$ S_i=\sum_{j\in \partial i}\tanh(\beta J_{ij})v_{j\to i}.$$
Then we set
positive entries of $S$ to $1$ and negative entries to $-1$, and zero entries to $1$ or $-1$ randomly. The 
obtained configuration should be correlated with the pattern, i.e. the overlap parameter which is
inner product of obtained configuration and a pattern
\begin{equation}\label{eq:ovl}
	O=\frac{1}{n}\sum_{i=1}^nS_{i}\xi_i^\mu,
\end{equation} 
is positive.

\begin{figure}
   \centering
    \includegraphics[width=0.6\columnwidth]{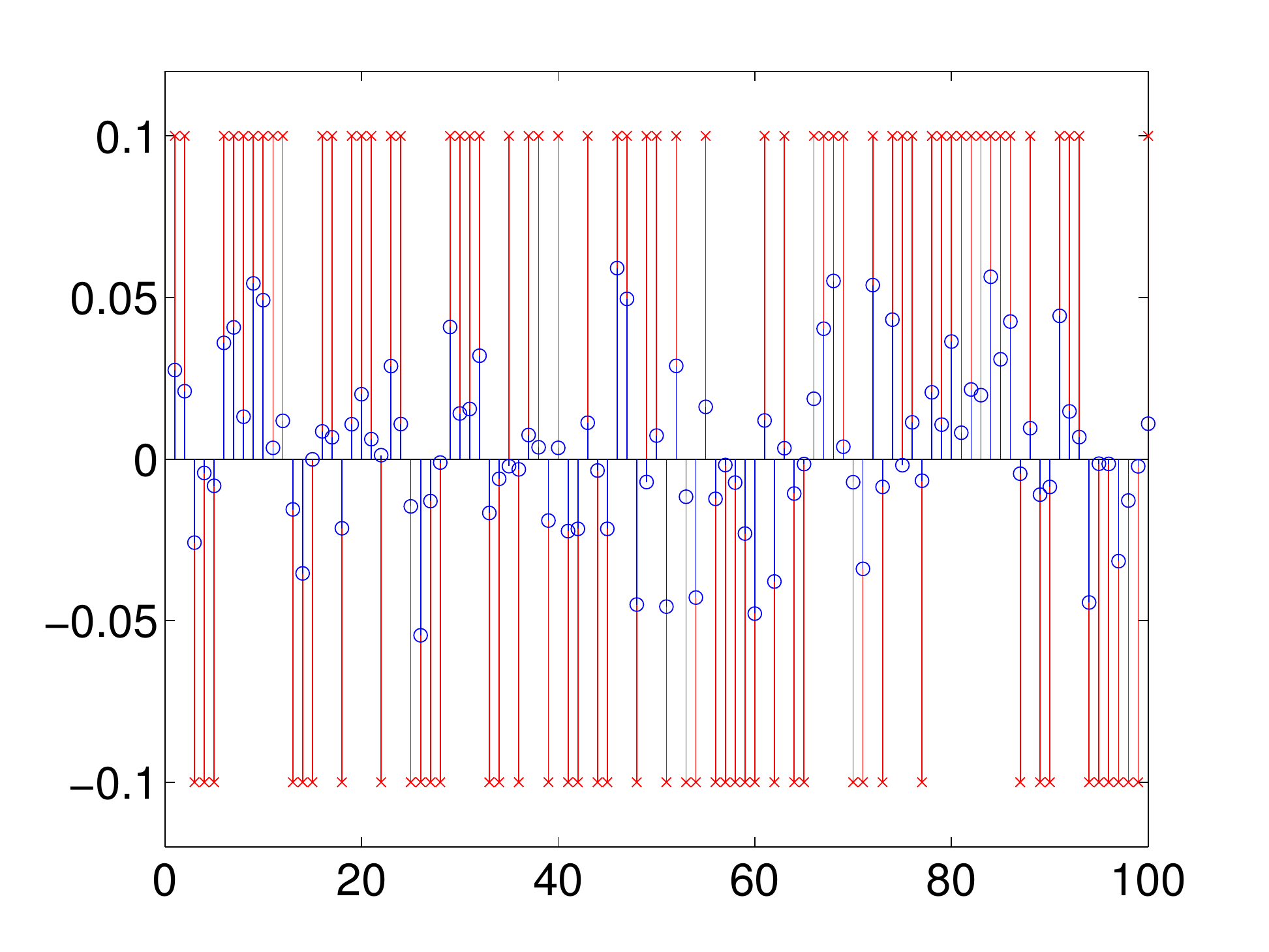} 
	\caption{
	Comparison of a binary pattern and one eigenvector of non-backtracking matrix.
	Red stars are components of the binary pattern and blue circles denote components of the eigenvector.
	Parameters are $n=100, \beta=0.02, P=3, c=28$.
	\label{fig:scatter_plot}}
\end{figure}

In Fig.~\ref{fig:scatter_plot} we plot one memorized binary pattern compared with one eigenvector of 
non-backtracking matrix, we can see that the sign of components of eigenvector are correlated with 
sign of the binary pattern.

\begin{figure}
   \centering
    \includegraphics[width=0.48\columnwidth]{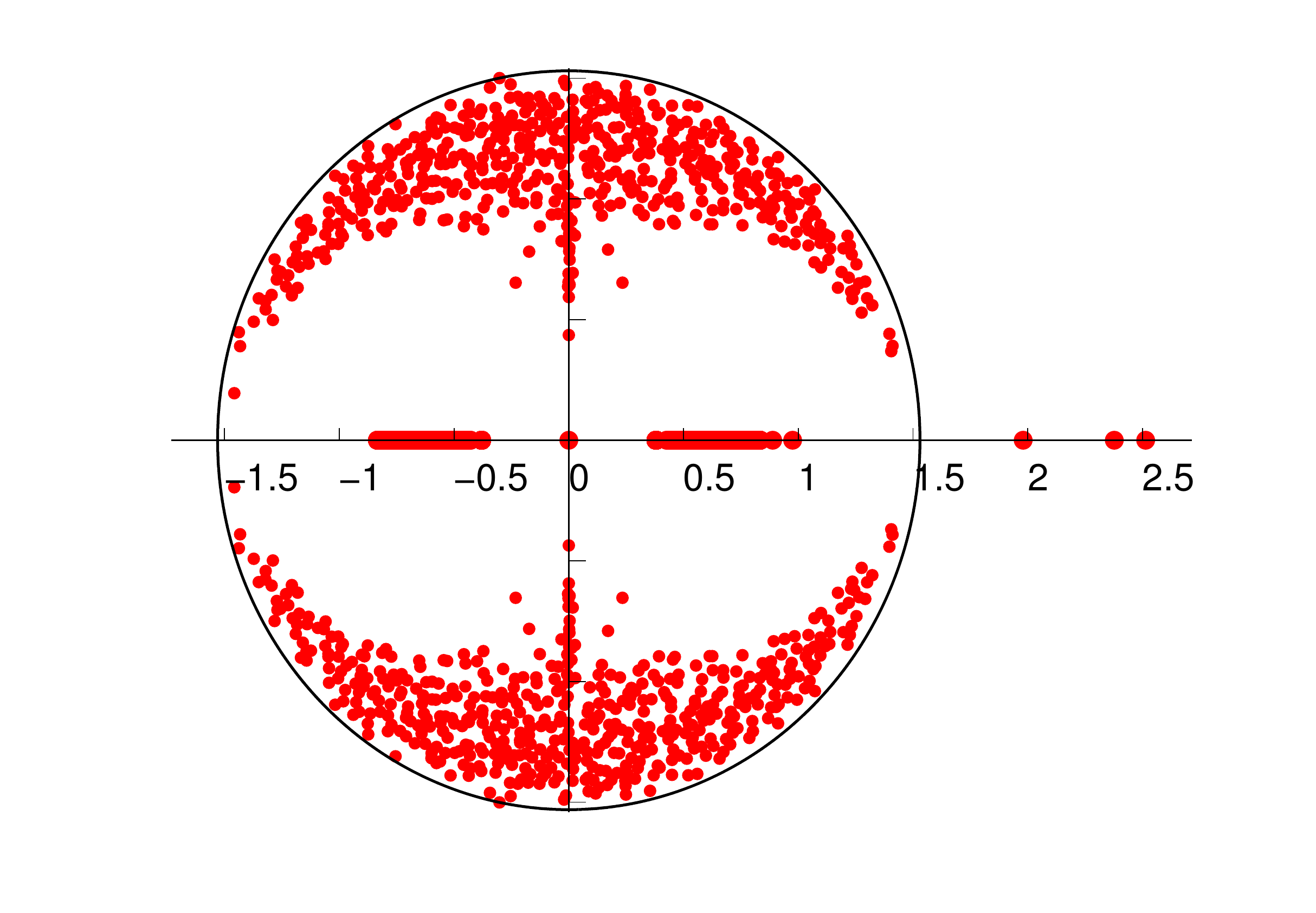} 
	\includegraphics[width=0.48\columnwidth]{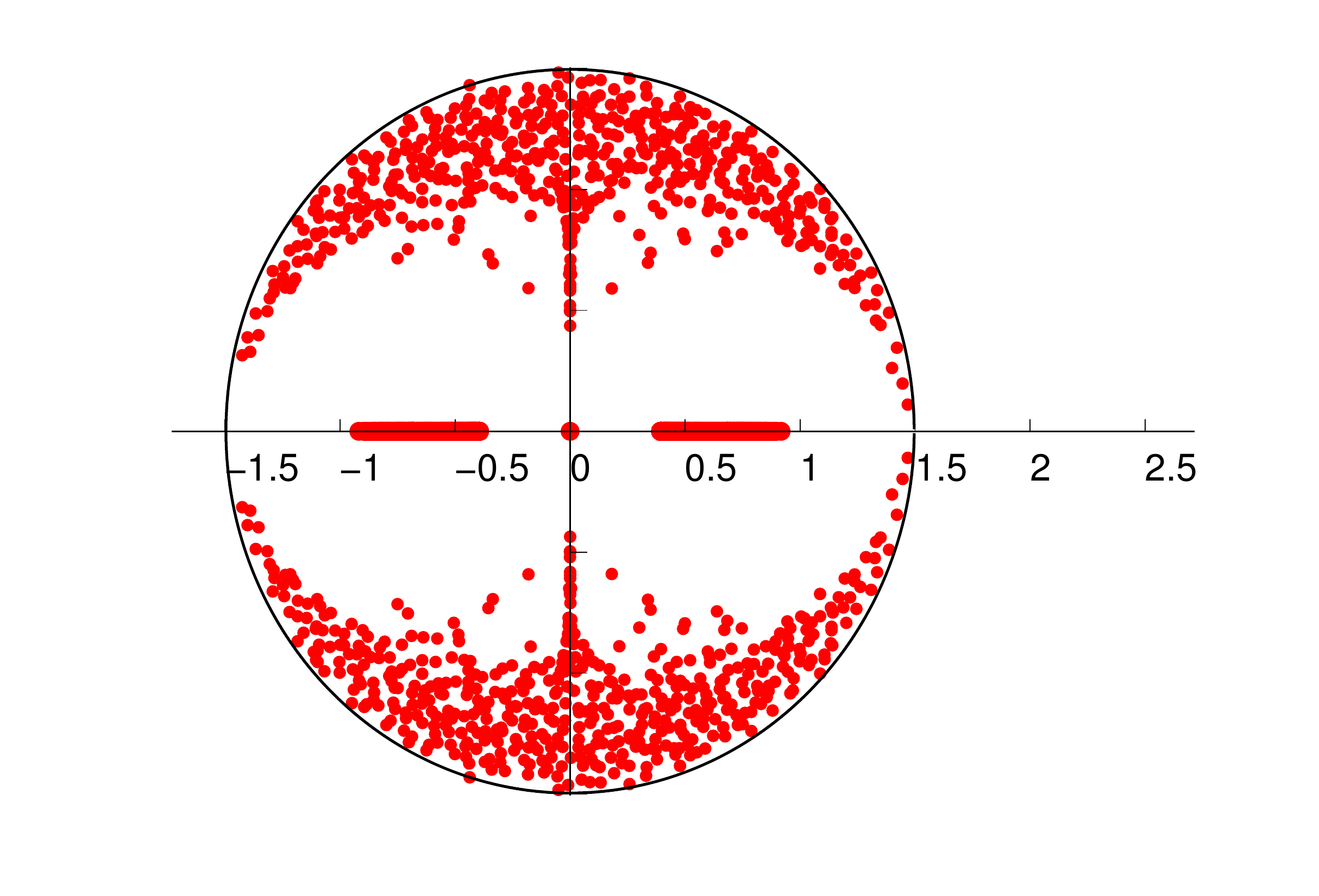}
	\caption{	
	Spectrum of non-backtracking operator for a Hopfield network with $n=500, c=8$ and 
	$3$ patterns with $\beta=1.2$ (left) and $12$ patterns with 
	$\beta=2.5$ (right).
	In left panel, three patterns are all memorized, overlap
    (Eq.\eqref{eq:ovl}) for three patterns obtained using first three eigenvectors 
	are $0.628$, $0.620$ and $0.448$.
	In right panel none of the patterns is memorized, system is in the spin glass phase.
	\label{fig:hop}}
\end{figure}

In Fig.~\ref{fig:hop} we plot the spectrum of a Hopfield network with different parameters
in the complex plane.
In left panel, $P=3$, three patterns are memorized, system is in retrieval phase. 
There are three eigenvectors outside the edge of bulk, and the associated eigenvectors are correlated 
with three stored patterns, with the overlap (Eq.\eqref{eq:ovl}) for three patterns much larger than 
$0$. In right panel, $P=12$, none of the patterns is memorized successfully. 
In spectrum, the edge of bulk is larger than $1$, there is no real eigenvalues outside the bulk, 
so system is in spin glass state.

In the thermodynamic limit, self-averaging property will make those $P$ eigenvalues identical. But in the 
finite-size networks, we could obtain different eigenvalues, telling us that $P$ patterns are not 
memorized equally strong. One example is shown in left panel of Fig.~\ref{fig:hop} where 
three real-eigenvalues outside the bulk are not equal to each other. 
When $P$ real-eigenvalues are all out of bulk but not equal, though 
from random initial vectors, iterating $C$ will always leads to the eigenvector associated with the 
largest eigenvalue, other patterns could be retrievable by other methods,
E.g. by running Glauber dynamics starting from a configuration or running BP from a 
random messages.
We have tested that by running Glauber dynamics we can reach other patterns according to the eigenvalues 
outside the bulk from random initial configurations. We think the pattern with the largest eigenvalue 
are indeed stronger than other two patterns, and has larger basin of attraction than other two 
patterns; however its basin attraction does not dominate the entire configuration space, so other two 
patterns are still able to attract Glauber dynamics.

Another situation is that not all $P$ eigenvalues are outside the bulk and larger than $1$. 
This means not all the 
patterns are memorized stably.  We have 
tested that in this case only patterns according to eigenvalues outside the bulk are attractive to Glauber 
dynamics. For those patterns according to eigenvalues inside bulk or smaller than $1$, Glauber dynamics 
starting from exactly the pattern will still run away and converge to configurations around other patterns. 
So those patterns are indeed not retrievable.

So as described above, we can find all real-eigenvalues and 
use all real-eigenvectors associated with those eigenvalues to retrieve pattern
simultaneously, as opposed to BP or Glauber dynamics, where only one pattern is 
retrieved at one time, without knowing neither total number of patterns, nor which patterns are
memorized stronger than the others.

Besides the phase diagram, another property that has been studied extensively in neural networks is the 
capacity of the network that we denote by $\mathcal {C}$. 
Capacity is the maximum number of patterns (divided by number of neurons) 
that could be stored successfully. 
This property is closely related to the strength of the memorized patterns. In our picture of 
spectrum of non-backtracking operator, by increasing number of patterns, number of real eigenvalues 
outside of bulk increases, but each of eigenvalue becomes closer to the bulk. When $P$ exceeds $\mathcal C$,
real-eigenvalues disappear and become complex eigenvalues, then system enters spin glass phase and losses 
all the memory.

A large amount of studies have been devoted to increase capacity and the retrieval abilities 
of neural networks by optimizing 
the learning rule or by optimizing the topology of the network \cite{anthony2009,mcgraw2003,kim2004,
morelli2004,zhang2008a,zhang2008,Braunstein2011}. In the following text of the paper we are going to 
discuss more on optimizing the topology of the network. We know that fully-connected Hopfield network can 
store number of patterns proportional to number of neurons, but the fully-connected topology is not 
biological realistic. On the other hand, when we dilute the network, i.e. by making the network more and 
more sparse by removing edges, number of patterns stored is decreasing but the number of patterns 
per edge is increasing \cite{Wemmenhove2003}. Thus one interesting question would be, 
given a network, how to make 
the network more sparse while keeping information of pattern stored by the network unchanged.
This problem could be studied easily by non-backtracking matrix,
since real eigenvalues outside the bulk tell us the stability of patterns, if we can keep those eigenvalues
unchanged in manipulating topology of the network, we can keep those patterns stable in the network.
A nature idea to keep eigenvalues could be by converting the problem to an optimization problem 
that at each time we remove the edge that decrease least the gap of eigenvalues,
$$\Delta =\sum_i(\lambda_i-|\lambda_{e}|),$$
where $\lambda_i$ is the eigenvalue corresponding to the pattern we want to keep, and $\lambda_{e}$ 
is the edge of bulk of non-backtracking matrix.
This process is easy to implement but the computation is time consuming because
selecting the edge that decreases least the spectral gap we have to remove every edge from the graph one
by one, compute eigenvalues, then add the edge back.
One way to make the process faster is to remove the edge that changes least the spectral gap among 
small randomly selected subset of edges.
Actually if we are interested in keeping one eigenvalue $\lambda$ instead of keeping the gap, 
we can do things much easier.

Assume after one edge removed, matrix $C$ becomes 
$$C_{\textrm{new}}=C+\Delta C,$$
If we look $\Delta C$ as a perturbation to the matrix $C$ and assume that after the perturbation, 
an eigenvalue changes from $\lambda$ to $\Delta \lambda$, 
a left eigenvalue changes from $u$ to $u+\Delta u$, and 
a right eigenvector changes from $v$ to $v+\Delta v$, then we have the following relation
$$(C+\Delta C)(v+\Delta v)=(\lambda + \Delta \lambda)(v+\Delta v).$$
Ignoring second order terms, we have
$$\Delta C v+C  \Delta v = \lambda\Delta v +\Delta \lambda v.$$ 
Multiplying left eigenvector $u$ in both sides of last equation results to
$$u\Delta C v+uC\Delta v=\lambda u\Delta v +\Delta \lambda u v,$$
then the expression of change of $\lambda$ by perturbation is written as:
\begin{equation}
	\Delta\lambda = \frac{u\Delta C v}{uv}.
\end{equation}
In our problem $\Delta C$ is actually a matrix with entries zero except those with 
edges $i\to j$ and $j\to i$ involved. Then the last equation can be written as
\begin{eqnarray}
	\Delta \lambda &=& \frac{1}{uv}
	\sum_{i\to j,k\to l}u_{i\to j}\Delta C_{i\to j, k\to l}v_{k\to l}\nonumber\\
	&=&\frac{1}{uv}\left [ u_{i\to j}\sum_{k\in\partial i\backslash j}\tanh(\beta J_{ik})v_{k\to i}
+v_{i\to j}\sum_{l\in\partial j\backslash i}\tanh(\beta J_{jl})u_{j\to l}\right. \nonumber\\
	&&+u_{j\to i}\sum_{l\in\partial j\backslash i}\tanh(\beta J_{jl})v_{l\to j}
	+ v_{j\to i}\left.  \sum_{k\in\partial i\backslash j}\tanh(\beta J_{ik})u_{i\to k}\right]
	\nonumber\\
	&=&\frac{2\lambda}{uv}(u_{i\to j}v_{i\to j}+u_{j\to i}v_{j\to i}).
\end{eqnarray}
So the change of eigenvalues becomes a function of left and right eigenvectors.
By computing once the eigenvectors, we know exactly which edge could change least the eigenvalue.
So we can make network more sparse while keeping information of a pattern stable,
by removing edges that give least $u_{i\to j}v_{i\to j}+u_{j\to i}v_{j\to i}$ iteratively. 
For more detail, at each step, left and right eigenvectors associated with 
a real eigenvalue are computed, one (or several edges) that 
minimizes $u_{i\to j}v_{i\to j}+u_{j\to i}v_{j\to i}$ are removed, then the above process are repeated 
until network is sparse enough.
Note this procedure is similar to the 
decimation algorithm using marginals of a message passing algorithm in solving constraint satisfaction 
problems \cite{Mezard2002}, where nodes having most biased marginals are removed (fixed) at 
each iterating. 

\begin{figure}
   \centering
    \includegraphics[width=0.6\columnwidth]{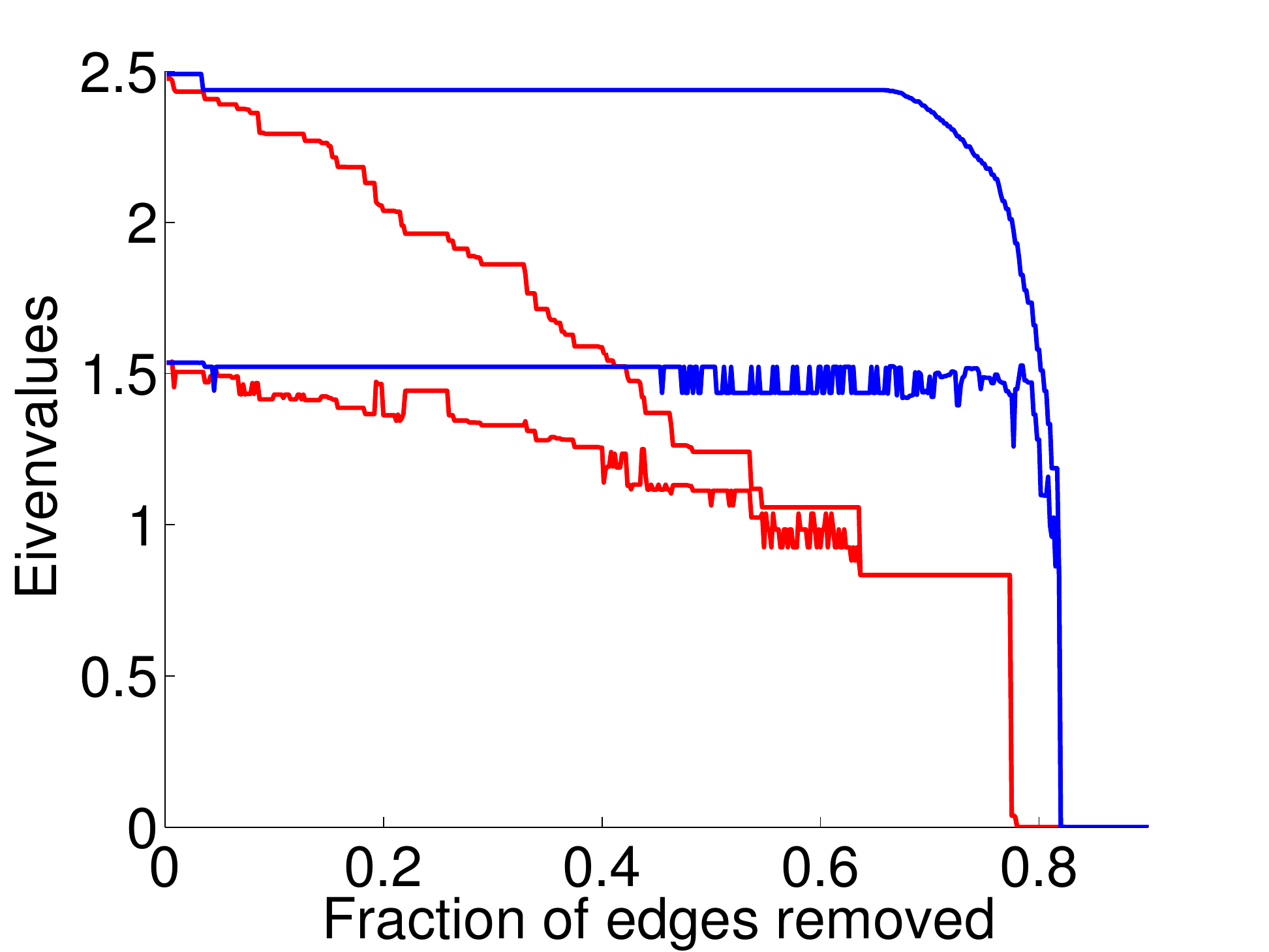} 
	\caption{	
	The largest eigenvalue and edge of bulk of a Hopfield network as a function of fraction of 
	edge removed from the network.
	Red lines denote random removing and blue lines denote our approach. In 
	both random removing and our approach, upper line represents the largest eigenvalue and the lower line 
	represents edge of bulk, that is the absolute value of the first complex eigenvalue.
	In this network  $n=200, c=6, P=2, \beta=1.2$.\label{fig:deci}}
\end{figure}

We did some numerical experiments on the edge-removing scheme and compared its performance with random
removing in Fig~\ref{fig:deci}, where the real eigenvalue and edge of bulk are plotted as 
a function of fraction of edges removed. We can see 
from the figure that random removing makes the real eigenvalue decreasing very fast while our process keep 
the eigenvalue unchanged up to more than half edges removed. 
Though edge of bulk in our process decreases also slower, the 
real eigenvalue joins bulk later than that in random removing.

Note that the problem of removing edges is just a simple example of optimizing topology of a neural network 
using the non-backtracking operator. Since information of stability of patterns can be obtained 
at same time by non-backtracking matrix we believe other kinds of topology optimization on neural networks 
can also be done easily using non-backtracking matrix.

\section{Conclusion and Discussions}\label{sec:dis}
As a conclusion, we have introduced non-backtracking operator for Ising model with general 
coupling distributions, it can be seen as a generalization of non-backtracking operator from a 
unweighted graph to weighted graph with an edge weighted by $\tanh\bra{\beta J_{ij}}$. 
In thermodynamic limit its spectrum gives us phase transitions of the model. And in 
single instances its eigenvectors can be used to control the model.

All above studies were made for Ising model without external fields. With external fields present, 
it has been shown \cite{parisi2014} that paramagnetic solution still exists but with finite magnetization. It would be 
interesting to extend current study to that case.

If one compares the spectrum of non-backtracking operator of a Hopfield model with $P$ patterns and that of 
a network generated by stochastic block model with $P$ communities, one find that they are quite similar. However the 
stochastic block model corresponds to a Potts model with $P$ states and one basin of attraction, 
as opposed to Hopfield model which corresponds to Ising model with $2$ states and $P$ basin of
attractions. So it would be interesting to study such connection in detail in future work.

Note that spectrum density of non-backtracking matrix of a graph has been computed in \cite{Saade14} using Belief Propagation. It 
would be interesting to compute non-backtracking matrix of Ising model using the same technique.

\begin{acknowledgements}
P.Z. was supported by AFOSR and DARPA under grant FA9550-12-1-0432.
We are grateful to Ruben Andrist, Florent Krzakala, Cristopher Moore, Alaa Saade and 
Lenka Zdeborov\'a for helpful conversations, and Federico Ricci-Tersenghi for discussing and pointing out
reference \cite{Mooij05,parisi2014}.
\end{acknowledgements}


\end{document}